\documentclass[conference]{IEEEtran}
\makeatletter
\def\ps@headings{%
\def\@oddhead{\mbox{}\scriptsize\rightmark \hfil \thepage}%
\def\@evenhead{\scriptsize\thepage \hfil \leftmark\mbox{}}%
\def\@oddfoot{}%
\def\@evenfoot{}}
\makeatother 
\IEEEoverridecommandlockouts

\usepackage{graphicx}
\usepackage{amsmath,amssymb}

\usepackage{algorithm}
\usepackage{algpseudocode}
\usepackage{amsmath}
\usepackage{graphics}
\usepackage{epsfig}
\usepackage{color}
\usepackage{subfigure}

\newtheorem{remark}{\underline{Remark}}[section]
\usepackage[numbers,sort&compress]{natbib}
\usepackage{flafter}

\newcommand{\mv}[1]{\mbox{\boldmath{$ #1 $}}}
\begin{document}
%
\title{3D Trajectory Optimization for Secure UAV Communication with CoMP Reception}
\author{\IEEEauthorblockN{Jianping~Yao\IEEEauthorrefmark{1}, Canhui~Zhong\IEEEauthorrefmark{1}, Zhihan~Liu\IEEEauthorrefmark{2}, and Jie~Xu\IEEEauthorrefmark{1}}
\IEEEauthorblockA{\IEEEauthorrefmark{1}School of Information Engineering, Guangdong University of Technology, Guangzhou 510006, China}
\IEEEauthorblockA{\IEEEauthorrefmark{2}Data Information Department, Guangzhou 510501, China}
\IEEEauthorblockA{E-mail: yaojp@gdut.edu.cn, zhongcanhui93@qq.com, liuzh\_gdut@qq.com, jiexu@gdut.edu.cn}
\vspace{-3em}
\thanks{J. Xu is the corresponding author.}
}

\maketitle
\setlength\abovedisplayskip{-0.01em}
\setlength\belowdisplayskip{-0.01em}

\begin{abstract}
This paper studies a secrecy unmanned aerial vehicle (UAV) communication system with coordinated multi-point (CoMP) reception, in which one UAV sends confidential messages to a set of distributed ground nodes (GNs) that can cooperate in signal detection, in the presence of several colluding suspicious eavesdroppers.
Different from prior works considering the two-dimensional (2D) horizontal trajectory design in the non-CoMP scenario, this paper additionally exploits the UAV's vertical trajectory (or altitude) control for further improving the secrecy communication performance with CoMP.
In particular, we jointly optimize the three dimensional (3D) trajectory and transmit power allocation of the UAV to maximize the average secrecy rate at GNs over a particular flight period, subject to the UAV's maximum flight speed and maximum transmit power constraints.
To solve the non-convex optimization problem, we propose an alternating-optimization-based approach, which optimizes the transmit power allocation and trajectory design in an alternating manner, by convex optimization and successive convex approximation (SCA), respectively.
Numerical results show that in the scenario with CoMP reception, our proposed 3D trajectory optimization significantly outperforms the conventional 2D horizontal trajectory design, by exploiting the additional degree of freedom in vertical trajectory.
\end{abstract}

\vspace{-0.5em}
\section{Introduction}
\vspace{-0.5em}
Recently, unmanned aerial vehicles (UAVs) are envisioned to play an important role in fifth-generation (5G) and beyond cellular networks.
On one hand, UAVs are emerged as a new type of aerial users that need to access cellular networks for enhanced communication quality and longer operation range \cite{ZengCellular2019}.
On the other hand, UAVs can also be employed as new wireless platforms (e.g., base stations, relays, and even wireless chargers) in the sky to provide flexible and on-demand wireless services to ground users, with improved transmission efficiency and better wireless coverage (see, e.g., \cite{ZengWireless2016,BerghLTE2016,XuUAV2018,WuCapacity2018} and the references therein).

However, due to the broadcast nature of wireless channels and the existence of strong line-of-sight (LoS) components over air-to-ground (A2G) links, UAV communications are more vulnerable to be eavesdropped by malicious ground nodes than conventional terrestrial communications.
Recently, physical layer security is becoming a viable new solution to protect UAV communications against malicious eavesdropping attacks (see, e.g., \cite{YaoSecure2016,YanSecret2018} and the references therein).
Different from conventional cryptology-based security technology, physical layer security is able to provide perfect security for wireless communication systems from an information theoretical perspective. Therefore, it is emerging and of great importance to conduct research on physical layer security in UAV communications.

In the literature, there have been various prior works investigating the integration of physical layer security in UAV communications and networks.
In general, these works can be roughly classified into two categories that considered the network-level performance analysis for large-scale random UAV networks via stochastic geometry \cite{Yao2019secrecy,ZhuSecrecy2018}, and the link-level performance optimization via UAV trajectory design and wireless resource allocation \cite{CuiRobust2018,ZhangSecuring2018,WangImproving2017,LiRobust2018,ZhouSecrecy2017,LeeUAV2018,ZhongSecure2018,CaiDual2018,ZhaoCaching2018}, respectively.
We are particularly interested in the link-level performance optimization in this paper.

Due to the high maneuverability, UAVs can properly control their locations over time (i.e., trajectories) as new design degrees of freedom for improving the secrecy communication performance (e.g., \cite{CuiRobust2018,ZhangSecuring2018,WangImproving2017,LiRobust2018,ZhouSecrecy2017,LeeUAV2018,ZhongSecure2018,CaiDual2018,ZhaoCaching2018}).
In particular, legitimate UAV transmitters can fly close to intended ground receivers to improve the legitimate channel quality and move far away from suspicious eavesdroppers to prevent the information leakage, thus improving the security of legitimate transmission \cite{CuiRobust2018,ZhangSecuring2018,WangImproving2017}.
Furthermore, UAVs can also be employed as aerial cooperative jammers that send artificial noise (AN) to confuse eavesdroppers, thus efficiently protecting the legitimate communications. For instance,
\cite{LiRobust2018,ZhouSecrecy2017} proposed to use a friendly aerial UAV jammer transmitting AN to help secure the communications between two ground nodes (GNs).
\cite{LeeUAV2018,ZhongSecure2018,CaiDual2018,ZhaoCaching2018} utilized one cooperative UAV jammer to secure the legitimate communication of another UAV, in which the communicating and jamming UAVs jointly design their trajectories and power allocations for performance optimization.
Nevertheless, these existing works on secrecy UAV communications mainly focused on a point-to-point legitimate UAV communication scenario from one legitimate UAV to one GN in the presence of suspicious eavesdroppers, in which the UAV flies at a fixed altitude and only adjusts its two-dimensional (2D) horizontal trajectory for performance optimization.

\begin{figure}[!t]
\setlength{\abovecaptionskip}{-0.5em}
\vspace{-0.2cm}
\centering
\includegraphics[width=6cm]{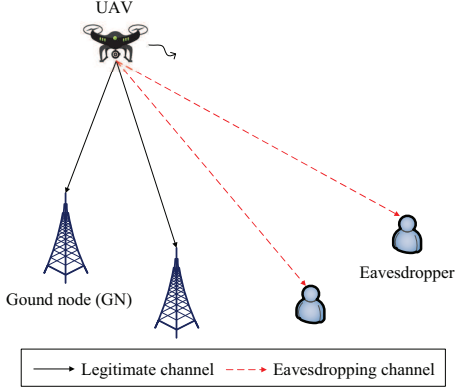}
\caption{Illustration of the secrecy UAV communication system with CoMP reception, in the presence of colluding suspicious eavesdroppers.}
\label{Fig:system model}
\end{figure}

Recently, coordinated multi-point (CoMP) transmission and reception has been recognized as a promising technique in wireless networks, which allows geographically distributed nodes (such as BSs) to cooperatively transmit or decode information at a symbol level.
By efficiently exploit the inter-cell interference via symbol-level signal processing, the CoMP technique can significantly enhance the communication reliability and increase the data-rate throughput, especially for users at the cell edge \cite{3GPPCoMP2011}.
The recently emerging cloud radio access networks (C-RAN) can be viewed as another form of CoMP \cite{CheckoCloud2015,SimeoneCloud2016}.
Motivated by its great success in cellular networks, there have been some prior works employing CoMP for secrecy communication \cite{bjornson2013optimal} and UAV communications \cite{LiuCoMP2019}, respectively.
For example, \cite{LiuCoMP2019} considered the CoMP transmission among a number of UAVs in the sky to serve ground users, in which these UAVs employ zero-forcing beamforming to cancel the interference among users for enhancing the throughput. However, to our best knowledge, there is no existing works investigating the secure UAV communications with CoMP. This thus motivates our investigation in this paper.

In this paper, we consider the secrecy UAV communication with CoMP reception, in which one UAV communicates with multiple legitimate GNs, in the presence of multiple suspicious eavesdroppers.
We consider that these GNs are enabled to cooperatively decode the legitimate messages sent from the UAV to defend against the eavesdropping attack, and these eavesdroppers are colluding in intercepting the messages, which corresponds to the worst-case eavesdroppers from the security perspective.
Different from previous works that focused on 2D horizontal trajectory design, we further exploit the vertical trajectory (or equivalently, altitude) via three-dimensional (3D) trajectory design, together with the transmit power allocation, to facilitate the secure communication.
In particular, our objective is to maximize the average secrecy rate from the UAV to the GNs over a finite communication period, by jointly optimizing the UAV's 3D trajectory and transmit power allocation, subject to the maximum flight speed and maximum power constraints.
Due to the coupling between transmit power and trajectory variables, the formulated secrecy rate maximization problem is non-convex, which is very difficult to be solved optimally.
Towards this end, we propose an alternating optimization based approach, which solves the transmit power allocation and trajectory design problems alternately, by using the convex optimization and successive convex approximation (SCA) techniques, respectively.
Numerical results show that the joint 3D trajectory and transmit power optimization greatly enhances the secrecy performance in the scenario with CoMP reception, as compared to other benchmark schemes with e.g. 2D trajectory optimization only.

\vspace{-0.5em}
\section{System Model}
\vspace{-0.5em}
As shown in Fig. \ref{Fig:system model}, we consider the secrecy UAV communication system, in which one UAV communicates to $K$ GNs in the presence of $J$ colluding eavesdroppers on the ground.
With CoMP reception, the distributed GNs can jointly decode the legitimate messages sent from the UAV.
We focus on a particular UAV mission/flight period with duration $T$ in second (s), which is discretized into $N$ time slots with equal duration $t_s\!=\!T/N$.
Let $\mathcal{N}\triangleq\{1, \ldots,N\}$ denote the set of slots.
Without loss of generality, we consider a 3D Cartesian coordinate system, in which the UAV exploits the fully-controllable mobility to change its 3D locations over time, in order to achieve better secrecy communication performance.
Let $(x[n],y[n],H[n])$ denote the time-varying 3D UAV location at time slot $n\in\mathcal{N}$, where $\mv{q}[n]=(x[n],y[n])$ denotes the horizontal location of the UAV, and $H[n]$ denotes its vertical location or altitude.
Also, suppose that $\mv{q}[0]=(x[0],y[0])$ and $\mv{q}[N+1]=(x[N+1],y[N+1])$ denote the UAV's pre-determined initial and final horizontal locations, and $H[0]$ and $H[N+1]$ denote the corresponding altitudes, respectively.
Let $\tilde{V}$ denote the UAV's maximum horizontal speed in meters/second (m/s), and $V=\tilde{V} t_s$ denote the maximum horizontal displacement of the UAV between two consecutive slots.
It then follows that
\begin{align}\label{Traj_constraint}
\|\mv{q}[n+1]-\mv{q}[n]\|\leq V, \forall n\in \{0\}\cup\mathcal{N},
\end{align}
where $\|\cdot\|$ denotes the Euclidean norm.
Similarly, let $\tilde{V}_{\text{up}}$ and $\tilde{V}_{\text{down}}$ denote the maximum vertical ascending speed and vertical descending speed, respectively.
Then, we have $V_{\text{up}}=\tilde{V}_{\text{up}} t_s$ and $V_{\text{down}}=\tilde{V}_{\text{down}} t_s$ as the maximum vertical ascending displacement and vertical descending displacement, respectively.
Let $H_{\text{min}}$ and $H_{\text{max}}$ denote the minimally and maximally allowed UAV flight altitudes, respectively, for e.g. safety reasons based on certain regulations.
Accordingly, we have
\vspace{-0.5em}
\begin{subequations}\label{Altitude_constraint}
\begin{align}
&H[n+1]-H[n]\leq V_{\text{up}}, \forall n\in \{0\}\cup\mathcal{N},\\
&H[n]-H[n+1]\leq V_{\text{down}}, \forall n\in \{0\}\cup\mathcal{N},\\
&H_{\text{min}}\leq H[n]\leq H_{\text{max}}, \forall n\in \mathcal{N}.
\end{align}
\end{subequations}
Furthermore, we denote the fixed horizontal location of GN $k$ as $\mv w_{k}\in\mathbb{R}^2$, $k\in\mathcal{K}\triangleq\{1, \ldots,K\}$, and that of eavesdropper $j$ as $\mv w_{ej}\in\mathbb{R}^2$, $j\in\mathcal{J}\triangleq\{1, \ldots,J\}$.

Different from prior works that normally considered the free-space path loss model for A2G links with path loss exponent $2$ \cite{CuiRobust2018,ZhangSecuring2018,LiRobust2018,WangImproving2017,LeeUAV2018,ZhongSecure2018,CaiDual2018,ZhaoCaching2018}, we adopt a more practical LoS channel model with a generic path loss exponent $\alpha$ that is determined by the radio propagation environment.
Accordingly, the channel power gain from the UAV to each GN $k\in\mathcal{K}$ at slot $n\in \mathcal{N}$ is given by
\begin{align}
\tilde{g}_{k}[n]=\frac{\beta_0}{d_{k}^{\alpha}[n]}=\frac{\beta_0}{\left(||\mv{q}[n]-\mv{w}_k||^2 + H^2[n]\right)^{\frac{\alpha}{2}}},
\end{align}
where $\beta_0$ denotes the channel power gain at the reference distance of $1$ m and $d_{k}[n]$ is the distance from the UAV to GN $k$ at slot $n$.
Similarly, the channel power gain from the UAV to eavesdropper $j\in\mathcal{J}$ at slot $n\in \mathcal{N}$ is
\begin{align}
\tilde{h}_{j}[n]=\frac{\beta_0}{d_{ej}^{\alpha}[n]}=\frac{\beta_0}{\left(||\mv{q}[n]-\mv{w}_{ej}||^2 + H^2[n]\right)^{\frac{\alpha}{2}}},
\end{align}
where $d_{ej}[n]$ is the distance from UAV to eavesdropper $j$ at slot $n$.

Next, we consider the secure communication from the UAV to the $K$ legitimate GNs.
At time slot $n\in \mathcal{N}$, let $p_t[n]$ denote the transmit power by the UAV and $s$ denote the transmitted symbol that is a circularly symmetric complex Gaussian (CSCG) random variable with zero mean and unit variance, i.e., $s \sim \mathcal{CN}(0,1)$.
In this case, the received signals at each GN $k\in\mathcal{K}$ and eavesdropper $j\in\mathcal{J}$ at slot $n\in \mathcal{N}$ are respectively given as
\begin{align}
y_{b,k}[n] =\sqrt{\tilde{g}_{k}[n]p_t[n]}s+n_{b,k},\\
y_{e,j}[n] =\sqrt{\tilde{h}_{j}[n]p_t[n]}s+n_{e,j},
\end{align}
where $n_{b,k}$ and $n_{e,j}$ denote the additive white Gaussian noise (AWGN) at the receivers of GN $k$ and eavesdropper $j$, respectively, each with zero mean and variance $\sigma^2$, i.e., ${n_{b,k}}\sim\mathcal{CN}(0,\sigma^2)$, ${n_{e,j}}\sim\mathcal{CN}(0,\sigma^2)$.
Then, the received signal-to-noise ratios (SNRs) from the UAV to GN $k\in\mathcal{K}$ and eavesdropper $j\in\mathcal{J}$ at slot $n\in\mathcal{N}$ are respectively given as
\vspace{-0.8em}
\begin{align}
\gamma_{k}[n]
=g_{k}[n]p_t[n],\\
\varepsilon_{j}[n]={h_{j}[n]p_t[n]},
\end{align}
where for notational convenience, $g_{k}[n] = \tilde{g}_{k}[n]/\sigma^2$ and $h_{j}[n] = \tilde{h}_{j}[n]/\sigma^2$ are defined as the channel-power-to-noise ratios from the UAV to GN $k$ and eavesdropper $j$, respectively.
Furthermore, suppose that the UAV is subject to a maximum average power $P_\text{ave}$ and a maximum peak power $P_\text{peak}$, where $P_\text{ave} \le P_\text{peak}$.
Then we have
\begin{subequations}\label{Power_constraint}
\begin{align}
&\frac{1}{N}\sum_{n\in\mathcal{N}} p_{t}[n] \le P_\text{ave}, \label{Power_ave}\\
&0\le p_{t}[n] \le P_\text{peak}, \forall n\in\mathcal{N}. \label{Power_peak}
\end{align}
\end{subequations}

By employing the optimal maximal ratio combining (MRC), the $K$ distributed GNs cooperate in decoding the legitimate message $s$ from the UAV.
In this case, the received SNR at the GNs at slot $n$ is given as
\begin{align}
\gamma[n] =\sum\limits_{k\in\mathcal{K}}\gamma_{k}[n] = \sum\limits_{k\in\mathcal{K}}g_{k}[n]p_t[n].
\end{align}
In addition, as the eavesdroppers are colluding, they can intercept/decode the confidential message $s$ from the UAV by using the MRC. Accordingly, the received SNR at the eavesdroppers is
\begin{align}
\varepsilon[n]=\sum\limits_{j\in\mathcal{J}}\varepsilon_{j}[n]=\sum\limits_{j\in\mathcal{J}}{h_{j}[n]p_t[n]}.
\end{align}

As a result, the secrecy rate from the UAV to the $K$ GNs at time slot $n$ (in bits-per-second-per-Hertz, bps/Hz) is given by \cite{ZhongSecure2018}
\begin{align}
\nonumber R[n]&=\left[\log_2\left(1+\gamma[n]\right)-\log_2\left(1+\varepsilon[n]\right)\right]^+,\\
&=\!\bigg[\!\log_2\!\bigg(\!1\!+\!\!\sum\limits_{k\in\mathcal{K}}\gamma_{k}[n]\!\bigg)\!\!-\!\log_2\!\bigg(\!1\!+\!\!\sum\limits_{j\in\mathcal{J}}\varepsilon_{j}[n]\!\bigg)\!\bigg]^+,
\end{align}
where $[x]^+\triangleq \max(x,0)$.

Our objective is to jointly optimize the 3D UAV trajectory $\{\mv{q}[n], H[n]\}$ and the transmit power allocation $\{p_t[n]\}$, to maximize the average secrecy rate from the UAV to GNs over the whole duration-$T$ period (i.e., $\frac{1}{N}\sum_{n\in \mathcal{N}}R[n])$, subject to the maximum UAV speed constraints in (\ref{Traj_constraint}) and (\ref{Altitude_constraint}), and the maximum power constraints in (\ref{Power_constraint}). The secrecy rate maximization problem is thus formulated as
\begin{align}
\nonumber(\text{P1}):\max \limits_{\{\mv{q}[n],H[n],p_t[n]\}}\frac{1}{N}\sum\limits_{n\in \mathcal{N}}R[n],\\
\nonumber ~\text{s.t.}~ (\ref{Traj_constraint}),~(\ref{Altitude_constraint}),~\text{and}~(\ref{Power_constraint}).
\end{align}
Note that the objective function of problem (P1) is non-smooth (due to the operator $[\cdot]^+$) and non-concave, with variables $\{p_t[n]\}$ and $\{\mv{q}[n], H[n]\}$ coupled.
Therefore, problem (P1) is a non-convex optimization problem, which is generally difficult to be optimally solved.

\vspace{-1em}
\section{Proposed Solution to Problem (P1)}
\vspace{-0.5em}
In this section, we propose an efficient solution to problem (P1).
First, we handle the non-smoothness of the objective function of problem (P1).
Towards this end, we omit the $[\cdot]^+$ operator in the objective function of (P1), and re-expressed (P1) as the following problem (P2).
According to Lemma 1 in \cite{ZhangSecuring2018}, the transmit power allocation optimization in (P1) can always lead to a non-negative secrecy rate at each time slot. Therefore, problems (P1) and (P2) are equivalent.
\begin{align}
\nonumber(\text{P2}): \mathop {\max }\limits_{\{ {\mv{q}}[n],H[n],p_t[n]\} } {\rm{ }}\frac{1}{N}\sum\limits_{n\in \mathcal{N}} {\bar R}[n],\\
\nonumber ~\text{s.t.}~ (\ref{Traj_constraint}),~(\ref{Altitude_constraint}),~\text{and}~(\ref{Power_constraint}),
\end{align}
where $\bar R[n] = \log_2\left(1+\gamma[n]\right)-\log_2\left(1+\varepsilon[n]\right), \forall n\in\mathcal{N}$.

Next, we focus on solving problem (P2), which, however, is still non-convex. To tackle this issue, we use the alternating optimization method to optimize the transmit power allocation $\{p_t[n]\}$ and UAV trajectory $\{\mv{q}[n], H[n]\}$ in an alternating manner, by considering the other to be given.

\vspace{-0.5em}
\subsection{Transmit Power Allocation Optimization}
\vspace{-0.5em}
First, we optimize the UAV's transmit power $\{p_t[n]\}$ under given UAV trajectory $\{\mv{q}[n], H[n]\}$, for which the optimization problem is expressed as
\vspace{-0.5em}
\begin{align}\label{subproblem_power}
\nonumber(\text{P3}):\mathop {\max }\limits_{\{p_t[n]\} } &\frac{1}{N}\sum\limits_{n\in \mathcal{N}} {\bar R}[n],\\
\nonumber ~\text{s.t.}&~ (\ref{Power_constraint}).
\end{align}
Notice that in this case, $\bar R[n]$ can be re-expressed as
\begin{align}
\bar R[n]= {\log _2}\left( {1 + {a_n}p_t[n]} \right) - {\log _2}\left( {1 + {b_n}p_t[n]} \right),
\end{align}
where $a_n$ and $b_n$ are constants given as
\begin{align}
&a_n=\sum\limits_{k\in\mathcal{K}}g_{k}[n],\\
&b_n=\sum\limits_{j\in\mathcal{J}}h_{j}[n].
\end{align}

It is evident that for problem (P3), under any time slot ${n\in \mathcal{N}}$, if the effective legitimate communication link has a stronger channel power gain than the effective eavesdropping link, i.e., $a_n > b_n$, then the rate function $\bar R[n]$ is a concave and monotonically increasing function with respect to the transmit power $p_t[n] \ge 0$; otherwise, it can be easily shown that the maximum of $\bar R[n]$ is zero, which is attained at $p_t[n] = 0$. Therefore, we only need to consider the power allocation optimization over a subset $\overline{\mathcal N}$ of time slots, with $\overline{\mathcal N} \subseteq {\mathcal N}$, and $a_n > b_n, \forall n\in \overline{\mathcal N}$. In this case, problem (P3) is equivalently re-expressed as
\begin{subequations}
\begin{align}
\nonumber (\text{P4}):\mathop {\max }\limits_{\{p_t[n]\} } &\frac{1}{N}\sum\limits_{n\in \overline{\mathcal N}} {\bar R}[n],\\
 ~\text{s.t.}~ &\frac{1}{N}\sum_{n\in\overline{\mathcal{N}}} p_{t}[n] \le P_\text{ave}, \label{P41}\\
 &0\le p_{t}[n] \le P_\text{peak}, \forall {n\in \overline{\mathcal N}}. \label{P42}
\end{align}
\end{subequations}

The reformulated problem (P4) is convex and thus can be optimally solved by the Karush-kuhn-Tucker (KKT) conditions \cite{boyd2004convex}.
Let $\upsilon$ denote the Lagrange multiplier associated with constraint (\ref{P41}), as well as $\underline\lambda_n$ and $\overline\lambda_n, {n\in \overline{\mathcal N}}$ denote the Lagrange multipliers associated with $p_t[n] \ge 0$ and  $p_t[n] \le P_\text{peak}$, respectively. Suppose that the optimal primal power allocation solution to problem (P4) are $\{p_t^*[n]\}$ and the optimal Lagrange multipliers are $\upsilon^*$, $\{\underline\lambda^*_n\}$, and $\{\overline\lambda_n^*\}$, respectively. Then based on the KKT conditions, they should satisfy the following sufficient and necessary conditions.
\begin{subequations}\label{KKT}
\begin{align}
\label{KKT1_1}&\upsilon^*\ge 0,\\
\label{KKT1_2}&0\le p_{t}^*[n] \le P_\text{peak}, \forall n\in\overline{\mathcal N},\\
\label{KKT1_3}&{\underline\lambda_n^*} \ge 0,~{\overline\lambda_n^*} \ge 0,~\forall n\in\overline{\mathcal N},\\
\label{KKT1_4}&{\underline\lambda_n^*} p_t^*[n] =  0,~\forall n\in\overline{\mathcal N},\\
\label{KKT1_4_2}&{\overline\lambda_n^*} (p_t^*[n]-P_\text{peak}) =  0,~\forall n\in\overline{\mathcal N},\\
\label{KKT1_5}&\frac{1}{{\ln 2}}\!\!\left(\!\! {\frac{{{a_n}}}{{1 \!+\! {a_n}p_t^*[n]}} \!-\! \frac{{{b_n}}}{{1 \!+\! {b_n}p_t^*[n]}}} \!\right) \!\!+\! {\underline\lambda _n^*}\!\!-\! {\overline\lambda_n^*}\!\!=\! \upsilon^*\!,\!\forall n\!\in\!\overline{\mathcal N}.
\end{align}
\end{subequations}
Based on the above KKT conditions in (\ref{KKT}), the optimal solution to problem ({P4}) is given as
\begin{align} \label{KKT_power}
p_t^*[n] = \min (P_\text{peak},~\tilde p_t^*[n]),~\forall n\in\overline{\mathcal N}.
\end{align}
where $\tilde p_t^*[n] = \bigg[ {\frac{{ - {{\rm{a}}_n} - {b_n} + \sqrt {{{({{\rm{a}}_n} - {b_n})}^2} + 4{{\rm{a}}_n}{b_n}(\frac{{{{\rm{a}}_n} - {b_n}}}{{\upsilon^* \ln 2}})} }}{{2{{\rm{a}}_n}{b_n}}}} \bigg]^+$.
By combining (\ref{KKT_power}) with the fact that the optimal $p_t[n]$ should be zero for any $n \in {\mathcal N} \setminus \overline{\mathcal N}$, it thus follows that the optimal solution to problem (P3) is
\begin{align}
p_t^\star[n] = \left\{ \begin{array}{l} p_t^*[n],~ n\in \overline{\mathcal N},\\
0,~~~~~~n \in {\mathcal N} \setminus  \overline{\mathcal N}.
\end{array} \right.
\end{align}
Notice that $p_t^*[n]$'s in (\ref{KKT_power}) are only dependent on the optimal Lagrange multiplier $\upsilon^* $, which can be obtained via a bisection search based on $\upsilon^*\left(\frac{1}{N}\sum_{n\in \overline{\mathcal N}} p_{t}^*[n] - P_\text{ave}\right)=0$.
Therefore, the transmit power allocation problem (P3) is finally solved optimally.

\vspace{-0.5em}
\subsection{Trajectory Optimization}
\vspace{-0.5em}
Next, we optimize the UAV trajectory $\{\mv{q}[n], H[n]\}$ under any given transmit power allocation $\{p_t[n]\}$, for which the optimization problem is expressed as
\begin{align}
\nonumber(\text{P5}):\mathop {\max }\limits_{\{ {\mv{q}}[n],H[n]\} } {\rm{ }}\frac{1}{N}\sum\limits_{n\in \mathcal{N}} {\bar R}[n],\\
\nonumber ~\text{s.t.}~ (\ref{Traj_constraint})~\text{and}~(\ref{Altitude_constraint}).
\end{align}

By introducing auxiliary variables $\{{\zeta}_k[n]\}$ and $\{{\eta}_j[n]\}$, problem (P5) is equivalently re-expressed as
\begin{subequations}
\begin{align}
&(\text{P6}):~\max\limits_{\{\mv{q}[n],H[n],{\zeta}_k[n],{\eta}_j[n]\}}~\frac{1}{N}\sum_{n\in \mathcal{N}}\hat{R}[n]\nonumber\\
\text{s.t.}~&{\zeta}_k[n]\!\geq\! \!\left(||\mv{q}[n]\!-\!\mv{w}_k||^2\!+\!H^2[n]\right)\!^{\frac{\alpha}{2}},\forall k\!\in\! \mathcal{K}, n\!\in\! \mathcal{N},\label{P23_1}\\
&{\eta}_j[n]\!\leq\! \!\left(||\mv{q}[n]\!-\!\mv{w}_{ej}||^2\!+\!H^2[n]\right)\!^{\frac{\alpha}{2}},\!\forall j\!\in\! \mathcal{J}, n\!\in\! \mathcal{N},\label{Nonconvex1}\\
&(\ref{Traj_constraint}),~(\ref{Altitude_constraint}),\nonumber
\end{align}
\end{subequations}
where $\hat{R}[n]\!=\!{\log _2}\left( {1 \!+\! \sum\limits_{k \in {\cal K}} {\frac{{{\beta _0}p_t[n]}}{{{\zeta _k}[n]}}} } \right) \!-\! {\log _2}\bigg( {1 \!+ \!\sum\limits_{j \in {\cal J}} {\frac{{{\beta _0}p_t[n]}}{{{\eta_j}[n]}}} } \bigg)$.
Notice that the function $\hat{R}[n]$ in the objective function and the right-hand-side (RHS) term in constraint (\ref{Nonconvex1}) are convex with respect to $\{\mv{q}[n], H[n]\}$, which make problem (P6) non-convex.
To tackle the non-convexity issue, we apply the SCA technique to obtain a converged solution in an alternating manner.
At each iteration $m\geq1$, suppose that the local trajectory point is given as $\{\mv q^{(m)}[n], H^{(m)}[n]\}$.
Then, we have the lower bounds for the function $\hat R[n]$ and the RHS term of (\ref{Nonconvex1}) as follows based on the first-order Taylor expansion.
\begin{align}\label{convex1}
&\left(||\mv{q}[n]-\mv{w}_{ej}||^2+H^{2}[n]\right)^{\alpha/2}
 \nonumber\\&\geq\alpha\Big[{H^{(m)}}[n](H[n]- {H^{(m)}}[n]) + ({{\mv{q}}^{(m)}}[n] - {{\mv{w}}_{ej}})
\nonumber\\\nonumber&\times ({\mv{q}}[n] - {{\mv{q}}^{(m)}}[n])\Big]\!\!\left(||\mv{q}^{(m)}[n]-\mv{w}_{ej}||^2+H^{(m)2}[n]\right)^{(\alpha-2)/2}\\
&+\left(||\mv{q}^{(m)}[n]-\mv{w}_{ej}||^2+H^{(m)2}[n]\right)^{\alpha/2}\triangleq E_{ej}^{\text{lb}}[n],
\end{align}
\begin{align}
\nonumber&\hat R[n]\!\ge\! {{\hat R}^{(m)}}\![n] \!\triangleq\! {\log _2}\!\bigg(\! {1 \!\!+\! \!\sum\limits_{k \in {\cal K}} \!{\frac{{{\beta _0}p_t[n]}}{{\zeta_k^{(m)}[n]}}} } \!\bigg)
\!\!-\!\! {\log _2}\!\!\bigg( \!{1 \!\!+ \!\!\sum\limits_{j \in {\cal J}} \!{\frac{{{\beta _0}p_t[n]}}{{{\eta_j}[n]}}} } \!\bigg)\\
 &-\frac{1}{\ln2} {{\!\bigg(\! {1 \!+\! \sum\limits_{k \in {\cal K}} {\frac{{{\beta _0}p_t[n]}}{{\zeta _k^{(m)}[n]}}} }\! \bigg)}^{\!\!-1}}\!\!\!\!\sum\limits_{k \in {\cal K}} {\!\bigg(\!{\frac{{{\beta _0}p_t[n]}}{{{{\zeta _k^{(m)2}[n]}}}}\left( {{\zeta _k}[n] \!-\! \zeta_k^{(m)}[n]} \right)}\!\! \bigg)}\label{convex2}.
\end{align}
Replacing (\ref{Nonconvex1}) and $\hat{R}[n]$ with the RHS terms in (\ref{convex1}) and (\ref{convex2}), respectively, problem (P6) is approximately expressed as the following convex optimization problem that can be efficiently solved by CVX \cite{grant2014cvx}.
\begin{align}
\nonumber(\text{P7}.m):~&\max\limits_{\{ {\mv{q}}[n],H[n],{\zeta_k}[n],{\eta_j}[n]\} } \frac{1}{N}\sum\limits_{n\in \mathcal{N}} {{{\hat R}^{(m)}}} [n],\\
\nonumber\text{s.t.}~~&{\eta}_j[n]\leq E_{ej}^{\text{lb}}[n],\forall j\in \mathcal{J}, n\in \mathcal{N},\\
&(\ref{Traj_constraint}),~(\ref{Altitude_constraint}),~\text{and}~(\ref{P23_1}).
\end{align}

Therefore, at iteration $(m+1)$, we update the UAV trajectory point $\{\mv{q}^{(m+1)}[n], H^{(m+1)}[n]\}$ as the optimal solution to the approximate problem (P7$.m$), under the local trajectory point $\{\mv{q}^{(m)}[n], H^{(m)}[n]\}$ in the previous iteration $m$.
As the iteration converges, we can obtain an efficient solution to problem (P5).

To sum up, we solve for the transmit power $\{p_t[n]\}$ and the trajectory $\{\mv{q}[n], H[n]\}$ in an alternating manner above, and accordingly, we obtain an efficient solution to problem (P2).
As the objective value of problem (P2) is monotonically non-decreasing after each iteration and the objective value of problem (P2) is finite, the proposed alternating optimization based approach is guaranteed to converge \cite{bertsekas1999nonlinear}. In Section IV, we will conduct simulations to show the effectiveness of the proposed algorithm.

\begin{remark}\label{remark1}
It should be noticed that the performance of our proposed alternating-optimization-based approach critically depends on the initial point for iteration. In this paper, we consider the following fly-hover-fly trajectory as the initial point. In this design, the UAV first flies straightly at the maximum speed from the initial location to the top of one GN, then hovers with the maximum duration, and finally flies straightly at the maximum speed to the final location.
We choose the hovering location as the point above the GN at the most central point among these GNs.
\end{remark}

\vspace{-0.5em}
\section{Numerical Results}
\vspace{-0.5em}
In this section, we conduct numerical results to validate the performance of our proposed design.
In the simulation, unless otherwise stated, we use the following settings to obtain the numerical results:
$\mv{w}_1 = (-100~\textrm{m},300~\textrm{m})$, $\mv{w}_2 = (0,300~\textrm{m})$, $\mv{w}_3 = (100~\textrm{m},300~\textrm{m})$,
$\mv{w}_{e1} = (-100~\textrm{m},100~\textrm{m})$, $\mv{w}_{e2} = (100~\textrm{m},100~\textrm{m})$,
$\mv{q}[0] = (-500~\textrm{m},0)$, $\mv{q}[N+1] = (500~\textrm{m},0)$,
$H[0] = H[N+1] = 200~\textrm{m}$, $H_\text{min}=150~\textrm{m}$, $H_\text{max}=250~\textrm{m}$,
$t_s=0.5~\textrm{s}$, $\alpha=2$, $\tilde{V}=25~\textrm{m/s}$, $\tilde{V}_{\text{up}}=4~\textrm{m/s}$, $\tilde{V}_{\text{down}}=6~\textrm{m/s}$,
$P_\text{ave}=30~\textrm{dBm}$, $P_\text{peak}=4P_\text{ave}$, and $\beta_0/\sigma^2=50~\textrm{dB}$.

\begin{figure}[!h]
\setlength{\abovecaptionskip}{-0.1em}
\vspace{-0.1cm}
\centering
  \includegraphics[width=6cm]{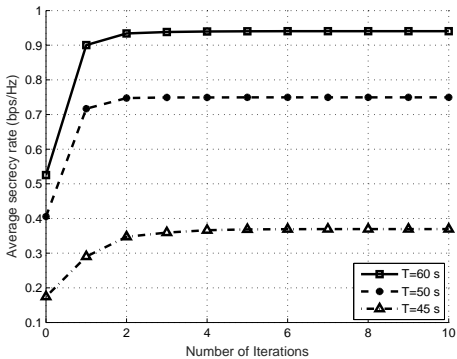}\\
  \caption{Convergence behavior of our proposed design.}
  \label{fig:coverage}
\end{figure}
Fig. \ref{fig:coverage} shows the convergence behavior of our proposed design, in terms of the average achievable secrecy rate versus the number of iterations.
It is observed that under different mission durations $T = 45$ s, $T = 50$ s, and $T = 60$ s, the average achievable secrecy rate converges within around $2$, $2$, and $4$ iterations, respectively. This thus validates the effectiveness and convergence of our propose design.

\begin{figure}[!h]
\setlength{\abovecaptionskip}{-0.1em}
\vspace{-0.05cm}
\centering
  \includegraphics[width=6cm]{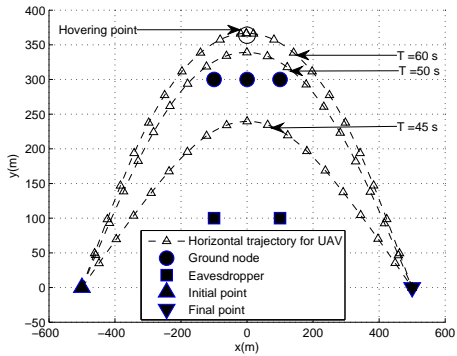}\\
  \caption{Obtained UAV horizontal trajectories by the proposed design, which are sampled every $2.5$ seconds.}
  \label{fig:trajectory}
\end{figure}
\begin{figure}[!h]
\setlength{\abovecaptionskip}{-0.1em}
\vspace{-0.1cm}
\centering
  \includegraphics[width=6cm]{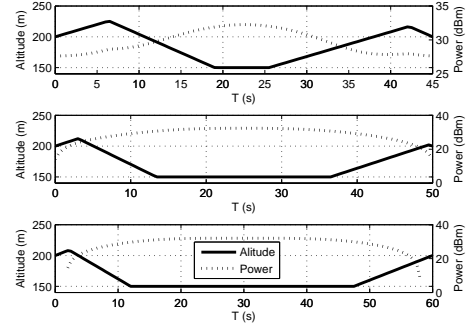}\\
  \caption{Obtained UAV altitude and transmit power over time by the proposed design.}
  \label{fig:altitude&power}
\end{figure}
Figs. \ref{fig:trajectory} and \ref{fig:altitude&power} show the obtained trajectories and transmit powers of the UAV by our proposed design, under mission duration $T = 45$ s, $T = 50$ s, and $T = 60$ s, respectively.
It is observed that for the three mission duration values, the UAV flies apart from eavesdroppers but close to the GNs following arc paths. When $T$ is large (e.g., $T =60$ s), the UAV is observed to hover at an optimized point with longest duration. By contrast, when $T$ is small (e.g., $T = 45$ s or $50$ s), the UAV is observed to fly at the maximum speed towards the hovering location as close as possible, but they cannot exactly reach there due to the time and speed limitations.
It is also observed that when the UAV is close to the eavesdroppers, it lifts its altitude and decreases the transmit power to prevent the undesirable information leakage.
By contrast, when UAV is far away from the eavesdroppers but close to the GNs, the UAV drops its altitude and increases the transmit power to enhance the desirable information transmission.
Under such an optimized trajectory and transmit power allocation policy, the UAV can significantly enhance the secrecy communication performance by effectively balancing between the desirable information transmission versus undesirable information leakage.

\begin{figure}[!h]
\vspace{-0.1cm}
\setlength{\abovecaptionskip}{-0.1em}
\centering
  \includegraphics[width=6cm]{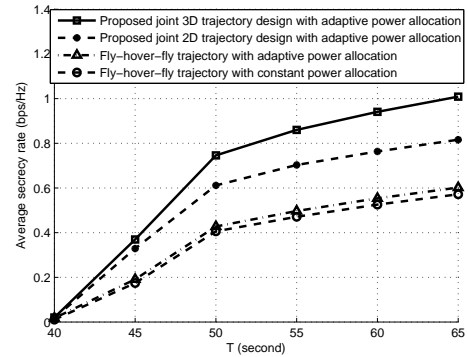}\\
  \caption{Average secrecy rate versus mission duration $T$.}
  \label{fig:rate}
\end{figure}
Fig. \ref{fig:rate} shows the average achievable secrecy rate versus the mission duration $T$.
For comparison, we consider the following three benchmark schemes.
\begin{enumerate}
\item Joint 2D trajectory design with adaptive power allocation: The UAV flies at a fixed altitude $H$, and jointly optimizes its 2D horizontal trajectory and the transmit power allocation to maximize the average secrecy rate. This design can be implemented by solving (P1) under the special case with  $H_\textrm{max} = H_\textrm{min} = H$. For this design, we set the fixed altitude as $H = 200$ m.

\item Fly-hover-fly trajectory with adaptive power allocation: The UAV adopts the fly-hover-fly trajectory in Remark \ref{remark1}, during which it adaptively optimizes the transmit power allocation by solving problem (P3).

\item Fly-hover-fly trajectory with constant power allocation: The UAV adopts the fly-hover-fly trajectory in Remark \ref{remark1}, during which it employs the constant power allocation, i.e., $p_t[n] = P_\textrm{ave}, \forall n\in\mathcal{N}$.
\end{enumerate}
It is observed that as $T$ increases, the secrecy rates achieved by all the four schemes increase. This is due to the fact that in this case, the UAV can fly closer towards the hovering location and hover there with longer duration, thus leading to higher average achievable secrecy rate.
When $T$ is small (e.g. $T = 40$ s), the two schemes with joint 3D and 2D trajectory design and adaptive power allocation are observed to have a similar performance as the fly-hover-fly trajectory with adaptive power allocation.
This is due to the fact that in this case, the block duration is only sufficient for the UAV to fly from the initial to final locations, and there is no additional time to adjust the trajectory for communication performance optimization.
When $T$ becomes large, the two schemes with joint 3D and 2D trajectory design and adaptive power allocation are observed to significantly outperform the two benchmark schemes with fly-hover-fly trajectory.
Furthermore, it is observed that there is a large performance gap between the proposed 3D trajectory design with adaptive power allocation versus the 2D trajectory design with adaptive power allocation.
This shows the significance of adapting the vertical UAV trajectory or altitude in enhancing the secrecy UAV communication performance with CoMP reception.

\vspace{-1em}
\section{Conclusion}
\vspace{-0.5em}
In this paper, we considered the CoMP reception-enabled secrecy UAV communication system, in which multiple GNs cooperatively detect the legitimate information sent from the UAV to enhance the legitimate communication performance.
We jointly optimized the UAV's 3D trajectory and transmit power allocation to maximize the average secrecy rate.
However, due to the non-convexity, this problem is generally difficult to be solved optimally.
Towards this end, we proposed to use the alternating optimization technique for solving the joint 3D trajectory and power allocation optimization, in an alternating manner, by convex optimization and SCA, respectively.
Numerical results showed that the proposed design significantly improves the average secrecy rate from the legitimate UAV transmitter to the GNs, as compared to other benchmark schemes.
How to extend the results to other scenarios, e.g., with multiple UAVs and multi-antenna GNs, are interesting directions worth further investigation.

\vspace{-1em}
\footnotesize
\bibliographystyle{IEEEtran}
\bibliography{myreference}
\end{document}